\documentclass[sigconf,authorversion]{acmart}

\AtBeginDocument{%
  }

\copyrightyear{2023}
\acmYear{2023}
\setcopyright{acmlicensed}\acmConference[ARES 2023]{The 18th International Conference on Availability, Reliability and Security}{August 29-September 1, 2023}{Benevento, Italy}
\acmBooktitle{The 18th International Conference on Availability, Reliability and Security (ARES 2023), August 29-September 1, 2023, Benevento, Italy}
\acmPrice{15.00}
\acmDOI{10.1145/3600160.3605005}
\acmISBN{979-8-4007-0772-8/23/08}

\usepackage{subcaption}
\usepackage{enumitem}
\usepackage{todonotes}

\begin{document}

\title[Tabletop As Social Engineering Prevention]{TASEP: A Collaborative Social Engineering Tabletop Role-Playing Game to Prevent Successful Social Engineering Attacks}

\author{Lukas Hafner}
\authornote{Both authors contributed equally to this research.}
\author{Florian Wutz}
\authornotemark[1]
\author{Daniela Pöhn}
\orcid{0000-0002-6373-3637}
\author{Wolfgang Hommel}
\email{[firstname.lastname]@unibw.de}
\affiliation{%
  \institution{Universität der Bundeswehr München, RI CODE}
  \city{Neubiberg}
  \country{Germany}
}

\renewcommand{\shortauthors}{Hafner and Wutz et al.}

\begin{abstract}
Data breaches resulting from targeted attacks against organizations, e.\,g., by advanced persistent threat groups, often involve social engineering (SE) as the initial attack vector before malicious software is used, e.\,g., for persistence, lateral movement, and data exfiltration. While technical security controls, such as the automated detection of phishing emails, can contribute to mitigating SE risks, raising awareness for SE attacks through education and motivation of personnel is an important building block to increasing an organization's resilience. To facilitate hands-on SE awareness training as one component of broader SE awareness campaigns, we created a SE tabletop game called Tabletop As Social Engineering Prevention (TASEP) in two editions for (a) small and medium enterprises and (b) large corporations, respectively. Its game design is inspired by Dungeons \& Dragons role-playing games and facilitates LEGO models of the in-game target organizations. Participants switch roles by playing a group of SE penetration testers and conducting a security audit guided by the game master. We evaluated the created game with different student groups, achieving highly immersive and flexible training, resulting in an entertaining way of learning about SE and raising awareness.
\end{abstract}

\begin{CCSXML}
<ccs2012>
   <concept>
       <concept_id>10002978.10003029</concept_id>
       <concept_desc>Security and privacy~Human and societal aspects of security and privacy</concept_desc>
       <concept_significance>500</concept_significance>
       </concept>
   <concept>
       <concept_id>10003120</concept_id>
       <concept_desc>Human-centered computing</concept_desc>
       <concept_significance>300</concept_significance>
       </concept>
 </ccs2012>
\end{CCSXML}

\ccsdesc[500]{Security and privacy~Human and societal aspects of security and privacy}
\ccsdesc[300]{Human-centered computing}

\keywords{Social engineering, gamification, tabletop, serious game, education, awareness}

\maketitle

\section{Introduction}

According to Costantino et al.~\cite{8417879}, 84\% of all cyber attacks conducted involving social engineering (SE) are successful. SE is an attack strategy that relies on the exploitation of human vulnerabilities through social manipulation by performing actions or leaking confidential information for fraudulent or malicious purposes against individuals or their organizations~\cite{9087851}. This may involve pretexting, phishing, and baiting, such as dropping malware-infested universal serial bus (USB) memory sticks that victims connect to their computers, e.\,g., due to curiosity. Many users are likely to click on links, download software, or provide information to people they trust, people with authority, and people with urgent requests. Social engineers thereby use this influence~\cite{WILLIAMS20181} within the SE attack cycle~\cite{kumar2015social}, involving the phases of information gathering, developing relationships, exploitation, and execution. In these phases, the adversaries identify the victims, gather information, and select the attack methods, before they engage with the victims, obtain the desired information, and then exit while removing their traces.

In order to make humans aware of SE, different mitigation strategies, such as endpoint security and awareness programs with training, posters, and information flyers, have been established~\cite{10.1145/3309074.3309083}. As technical solutions cannot detect and mitigate all attempts, individual awareness is of utmost importance. Learning defensive techniques can help, but changing the way a person thinks and making vulnerabilities visible may be a turn in defending against SE attacks. Future cyber security professionals must not only have the knowledge and skills to defend against all types of attacks, but must have an understanding of the psychological aspects of SE. Teaching SE in an information security class is possible in a lecture-based format. This has the drawback of not involving interactive hands-on training. Practically applying SE though needs to go through a rigid approval process with legal and ethical issues in most organizations, for good reasons. In order to provide a kind of hands-on awareness training in a controlled environment, games such as tabletops can be applied. As no similar game is known to the authors and training for SE is limited to Ngambeki et al.~\cite{9637328} approach of a real-world client penetration test, we designed and implemented Tabletop As Social Engineering Prevention (TASEP). TASEP is a tabletop role-playing game (TRPG) with two different LEGO settings (one small and medium-sized enterprise (SME) and one larger organization), three pre-defined scenarios each, and gameplay based on the principles of Dungeons \& Dragons (D\&D).

The contribution of this paper is as follows:
\begin{enumerate}
\item Design of tabletop models based on LEGO;
\item Concept of a game system with skills similar to D\&D for SE;
\item User studies conducted with both technical-savvy and non-technical-savvy players.
\end{enumerate}

The remainder of the paper is as follows: We first motivate our work by reviewing the state of the art in Section~\ref{sec:motivation}. Then, we outline our concept in Section~\ref{sec:concept} and its implementation in Section~\ref{sec:implementation}. TASEP has been practically evaluated, as summarized in Section~\ref{sec:evaluation} and is discussed in Section~\ref{sec:discussion}. Last but not least, we provide an outlook.

\section{Motivation}
\label{sec:motivation}

In this section, we motivate TASEP by reviewing various related work, ranging from SE to gamification and serious games for cyber security. Ngambeki et al.~\cite{9637328} designed a SE course based on literature and an overview of existing courses and tools. The proposed course consists of (a) the history, ethics, and law of SE; (b) the psychology of SE; (c) the tools of SE; (d) the prevention and mitigation of SE; and (e) course activities. The course activity consists of a theoretical case study and a group assignment with a real-world client. Dimkov et al.~\cite{10.1145/1920261.1920319} propose two methodologies for physical penetration testing using SE. Although a real-world exercise would be the most practical approach, it might not be possible. Hence, other ways to provide hands-on training are required.

Gamification describes the process of applying game-like elements and mechanics to non-game contexts, such as education. It can include features such as points, badges, leaderboards, challenges, and other incentives to encourage participation~\cite{10.1145/3167132.3167347}. Game-based learning is a method of active learning in digital or non-digital settings, whereas the variant of serious games is explicitly designed to apply these features for a specific purpose and audience with an educational purpose~\cite{abt1987serious}. Tabletop games are a broad category of games played on a flat surface. As pointed out by Chang et al.~\cite{10.1145/2669485.2669496}, there has been substantial research on interaction techniques for digital tabletop systems, including tangible boards, augmented reality~\cite{10.1145/3399715.3399836}, the integration of physical devices~\cite{4660178}, and immersion~\cite{murray1997hamlet}. In contrast, Farkas et al.~\cite{10.1145/3410404.3414224} analyze immersion for board games.

One well-known TRPG is D\&D. It is played with around five players and a dungeon master. Each player creates a character with abilities, characteristics, and resources. The character has an influence on the possible options. On the other hand, random decisions by dice roll and master decision include a certain level of uncertainty. Hence, the player is also dependent on their creativity to play with the options. The game master tells the story, which the players should follow, and may balance actions through master decisions. Thereby, they plan and orchestrate the games and partly apply creativity-supporting tools, as outlined by Tchernavskij et al.~\cite{10.1145/3527927.3532798}. Katifori et al.~\cite{10.1145/3555858.3555918} highlight the importance of creative aspects as motivation for the game master and the player group dynamics as key factors. The game D\&D is also applied in research. Veldthuis et al.~\cite{10.1145/3507923.3507927}, for example, use it for quests to engage computer science students with the goal to develop soft skills. Borodina et al.~\cite{10.1145/3337722.3342236} examine the players' perceptions of oddly shaped dices, whereas Plijnaer et al.~\cite{10.1145/3383668.3419931} enhance the game experience through tangible data visualization.

While several authors explore gamification for information security training (see, e.\,g., DeCusatis et al.~\cite{10.1145/3548771.3561409}, Gutfleisch et al.~\cite{9794181}, and Capture the Flag competitions), only a few authors focus on SE. Sharif and Ameen~\cite{9664771} provide a review of gamification for information security training. Denning et al.~\cite{10.1145/2508859.2516753} design and evaluate the card game Control-Alt-Hack to educate security awareness. Dixon et al.~\cite{10.1145/3290607.3313026} examine the possibility of cyber security educational games for phishing. The authors find a disconnect between casual and serious gamers, where casual gamers prefer simple games in comparison to serious gamers with a congruent narrative or storyline. Reinheimer et al.~\cite{10.5555/3488905.3488920} investigate phishing awareness and education over time. Videos and interactive examples performed best in the study. Mink and Freiling~\cite{10.1145/1231047.1231056} propose that teaching offensive methods yields better security professionals than teaching defensive techniques alone. Fabula Games~\cite{fabula} designed several mini-games concerning phishing, baiting, and tailgating. Beckers and Pape~\cite{7765507} propose a card game with the goal of identifying weaknesses in oneself and the organization based on escape plans. A similar strategy is applied by the game The Social Engineer~\cite{10.1145/3383668.3419917}, where a player has to conduct a penetration test in a cyber security company in a virtual reality environment.

\emph{Summary:} Although serious games are increasingly being used to teach specific topics and make users aware of cyber security, there is a lack of games related to SE. The aim of our approach is to create a dynamic game environment in which the players plan and carry out specific attacks in a realistic abstraction of an SME or large organization. Through this, their actions should raise awareness.

\section{Concept of TASEP}
\label{sec:concept}

In the following, we describe the underlying didactic concept, the developed models, and the designed gameplay.

\subsection{Methodology and Didactic Aspects}

Based on Benjamin Franklin's quote, ``Tell me and I forget. Teach me and I remember. Involve me and I learn.'', a realistic abstraction of the real world is needed to enable active learning. We design an interactive TRPG, in which players act as social engineerer. Thereby, we receive an abstraction of the real world, while at the same time, the participants (and the game master) are dealing with uncertainties and changing conditions through the gameplay. The reverse strategy enables the participants to be aware of possible attack vectors and defense mechanisms without pinpointing certain behaviors. We choose D\&D as it is a well-known TRPG as the first prominent game, which is also widely available in several languages. Hence, there is a likelihood that the players already know the game play. However, it is not necessary. Besides a theoretical introduction to the topic of social engineering and playing SE Tabletop, a debriefing with at least a discussion of countermeasures and awareness thereafter is required. Due to this concept, all neuro-didactic principles of Caine and Caine~\cite{Caine1990UnderstandingAB} and the COoperative Open Learning (COOL) principles~\cite{cool} are observed. During the game, the participants can see the physical elements, ranging from the building at the beginning to all floors at the end. This reflects the shift from digital attacks to the various equally physical possibilities and human elements of SE. The participants form groups to solve their missions, which require cooperation in teams, the acquisition of domain knowledge, critical thinking, and logical reasoning. The groups can be formed by, for example, existing knowledge, individual learning speed, and other characteristics. Thereby, SE Tabletop applies all stages of Bloom's taxonomy~\cite{forehand2005bloom} with an increasing level of difficulty. In addition, the participants acquire not only professional competencies and skills, but also social competencies, and they reflect on their behavior and possible consequences afterward.

\subsection{Models of TASEP}

The playing field should be as realistic as possible, yet simple enough to get started easily and focus on game progress without limiting creative problem-solving. In addition, it should be possible to modify the field during the play on short notice in order to simulate in-game events or a scenario change. Since most people we know have played with LEGO bricks in their youth and the concept is known, it was decided to use LEGO bricks to model the buildings and workspaces of the target organization in a modular setting. This means that at the beginning, the building is assembled as a whole, and only after exploration, each floor is placed separately, gradually removing the ``fog of war'' and allowing more insight. Until then, the players can only use the description provided by the game master. In addition, to have a uniform modular structure and comparable requirements, the floor plans of the individual scenarios (see Section~\ref{sec:scenarios}) should not differ.

As SMEs and big organizations vary due to their size as well as office equipment and security countermeasures, we designed different models with various scenarios but applied the same game principle. Thereby, the players can adapt their knowledge without a learning interruption caused by the introduction. According to the Dunbar number~\cite{DUNBAR1992469,dunbar}, individuals can stay in contact with around 150 persons. Hence, this number can be used to differentiate the size of an organization, where anonymity can be applied.

\subsubsection{SME}

To represent a startup company in the SME case, the media image of a micro-enterprise is applied. There are a few office rooms and floors that are rented in a house or office building complex. To increase the complexity, additional rooms or floors can be added. The structure of the floors themselves reflects the widespread morality of these companies, which is based on a more relaxed and family-based structure with short communication paths and fewer security restrictions. Hence, the workplaces are less isolated, a tea kitchen is provided instead of a canteen, and the interior architecture does not follow any design-technical stipulations. Several mitigation strategies from bigger organizations, such as a welcome desk and restricted areas, typically do not apply.

\subsubsection{Large Organization}

Large organizations have at least one building, sometimes even several buildings, with a fence surrounding the area (not taking different locations into account). Consequently, the following components are relevant:

\begin{itemize}
\item Parking lot, e.\,g., for USB drops;
\item Large paper garbage cans, e.\,g., for dumpster diving;
\item Smoking areas, tea kitchen, and areas for conversation;
\item Storage rooms to elucidate supply chains using crates;
\item Restricted areas;
\item Employees' computers with different permissions;
\item Server rooms;
\item Windows to observe the behavior from the outside;
\item Conference rooms, doors with secured access, and secretariats shielding executive offices.
\end{itemize}

Optional features include security guards, reception staff, cameras, and physical barriers. Since these components are the same for all possible game scenarios for large organizations, it is possible to use the same model with small modifications for all scenarios. This also saves time, as no complete new model is necessary.

\subsection{Gameplay}

Since D\&D is a well-known game, it serves as the basis for the concept of the gameplay and the scenarios.

\subsubsection{Outline of TASEP}

Similarly to several TRPGs, the game consists of several players and a game master (GM). The GM tells the story and interacts with the players to explain the setting. The GM also gives a voice to non-player characters, such as the employees of the corresponding scenario. Depending on the knowledge of the players, they can also act as lecturers. The social aspect of SE is covered by the interactions between the players and the GM during the scenarios. The players' freedom of choice is only affected by their characters and the setting of the scenarios. To define the individual strengths and weaknesses of a character, the players use skill points similar to the D\&D character sheet. In the current variant, the players will not be rewarded with skill points to improve their characters during the game. In order to represent the course of a SE attack, a 20-sided dice (D20) is applied. Such dice are often used in TRPGs in combination with skills to decide the outcome of a situation. Thereby, the players have to adapt to the current situation as a group, leading to the desired group dynamics. In addition, the GM can apply the D20 to introduce randomness for a decision~\cite{10.1145/3337722.3342236}.

\subsubsection{Scenarios}
\label{sec:scenarios}

The basic gameplay is specified by scenarios. These include the mission of the players, their background, and motivation, as well as available resources. In addition, the target objective is specified. These scenarios are based on real-world examples, but can be adapted to facilitate the creativity of the GM~\cite{10.1145/3555858.3555918}. In addition, these elements, together with the corresponding models, are relevant drivers for immersion~\cite{10.1145/3410404.3414224}. There is at least one intended solution for each scenario. However, due to the dynamics of the gameplay, it does not need to be followed. The first scenario per model is intended to act as an introduction to the game. For this purpose, a wide variety of attack options should be possible. Further scenarios should increase the level of difficulty.

The example of the IT security consulting organization TestArmy CyberForces is used~\cite{pizza}. TestArmy CyberForces acted as pentester for an international company. At the beginning of the pentest, employees received an email advertising a new pizza delivery service near the company. Motivated by a 30\% discount promotion for a limited time, a ``pizza day'' was organized and several boxes were ordered. The delivery service had its own homepage. The attackers bought the pizzas elsewhere and delivered them, using their own logo, to their victims. In addition, they distributed promotional gifts in the form of LEDs lights with USB ports. These changed their colors according to the rhythm of the music played. The surprised employees accepted the gifts immediately
by connecting them to the company computers. This gave the pentesters external access to the system.

\section{Implementation}
\label{sec:implementation}

In this section, we describe the implementation based on the skill system, skills, scenarios, and models.

\subsection{Skill System}

It is assumed that a master of their craft is successful about 90\% of the time, while a layman only 10\%. To implement these probabilities, a D20 is used in conjunction with a TRPG skill system, as described in the following section. A decision is successful when a threshold value is exceeded by the sum of the dice result and the numeric value of the associated skill. The player must roll a 19 or 20 to get a positive outcome, receiving the threshold of 18 points.

\begin{align}
    P(success) &= 10\% \\
    P(success) &= \frac{\text{\textit{No. (desired)}}}{\text{\textit{No. (possible)}}} \\
    10\% &= \frac{\text{\textit{No. (desired)}}}{20}\\
    \text{\textit{No. (desired)}} &= 2
\end{align}

Next, it is important to determine the necessary number of skill points in total. According to the principle that no one can do everything perfectly, the number of skill points should be realistic to encourage team play. Therefore, the \emph{proficiency} principle is introduced by rolling the dice twice for decision-making. The better result determines the outcome. Since the probabilities of failure of the two throws do not differ, they can be combined.

\begin{align}
    P(failure) &= P(failure\ 1st) * P(failure\ 2nd)\\
    10\% &= P(failure\ 1st) * P(failure\ 2nd)\\
    P(failure\ per\ throw)^{2} &= 10\%\\
    P(failure\ per\ throw) &= \sqrt{10\%} = 0.316
\end{align}

As a D20 moves in $0.05$ steps, the value is rounded up to $0.35$. This results in a desired failure chance of 35\% to exceed the previously determined threshold of 18 points.

\begin{align}
    P(failure) &= 35\%\\
    P(success) &= 65\%\\
    P(success) &= \frac{\text{\textit{No. (desired)}}}{\text{\textit{No. (possible)}}} \\
    \textit{No. (desired)} &= 65\%*20 = 13 \\
\end{align}

The threshold of seven points needs to be aligned with the original threshold, which then forms a unified threshold of eleven.

\begin{align}
  \textit{Threshold(unified)} &= \text{\textit{Threshold(original)}} - Threshold(new)\\
  \textit{Threshold(unified)} &= 18 - 7 = 11 
\end{align}

\subsection{Skills in TASEP}

Table~\ref{tab:skills} shows the intended superclasses and subclasses of skills, possible attacks, and possible inventory, if the proficiency level is reached. This selection is based on the relevant skills required for various SE approaches.

\subsubsection{Overview of the Skills}

\emph{Social skills} are mainly about interpersonal interactions. Manipulation aims at the communicative level, using methods such as persuasion or intimidation, to achieve its goals. For example, conversations can be steered in a certain direction, or sensitive information can be obtained through blackmail.
Appearance stands for credibility and charisma, i.\,e., how trustworthy the character appears. In this way, an unsuccessful attempt at manipulation can be compensated for by a strong appearance. A real-life example of this is the CEO fraud. The ability to improvise and expert disguise are summarized under adaptation. A master of adaptation could play external tech support, as they have the necessary appearance to appear believable.

\emph{Technical skills} contain the relevant skills for the computer-based SE approaches. Design combines both web and graphic design. For example, a character is able to create the custom logos and websites of a fake organization. Forging makes it possible to copy electronic key cards or conventional keys. A programmer enables players to create malware for USB drops or scareware or to perform attacks such as Domain Name System (DNS) spoofing. Analysis describes how well masses of information can be dealt with, such as after a successful dumpster diving attack. On the other hand, such a character can do network analysis to gain the knowledge needed to penetrate a large corporate network.

\emph{Finesse} describes the character's physical skills and sensitivity. Lock picking enables the character to enter locked rooms or sections of buildings. In addition, the contents of locked filing cabinets or similar furniture can be accessed. Fitness is an ability that allows the character to perform physically demanding acts, such as climbing fences. Sleight of hand covers common tricks used by pickpockets. This ranges from the traditional theft of key cards or wallets to the undetected theft of USB sticks from desks.

\begin{table*}[!htbp]
\centering
\caption{Skill categories}
\label{tab:skills}
\begin{tabular}{llll}
\toprule
\textbf{Super-category} & \textbf{Sub-category} & \textbf{Attack} & \textbf{Inventory for proficiency} \\ \midrule
Social skills & Manipulation & Quid pro quo & Bribes \\
                       & Appearance         & Shoulder surfing, pretexting & Fake IDs  \\
                       & Adaptation & Tailgating, pretexting &  Costumes \\ \midrule
Technical skills & Design & Scareware & Printer, server\\
                       & Forging & Badge surveillance & Key card copier\\
                       & Programming & Scareware, DNS spoofing, watering hole & Custom IT gadget \\
                       & Analysis & DNS spoofing, dumpster diving & Forensic equipment \\ \midrule
Finesse & Lock picking & Opening doors & Burglar tools \\
                       & Fitness & Sneaking & Smuggling of items \\
                       & Sleight of hand & Theft & Distraction \\ \bottomrule
\end{tabular}
\end{table*}

Specific attacks require two skills with separate random decisions for a positive result. Pretexting needs, e.\,g., appearance and adaptation. In addition, specific SE attacks may have two actions. For example, dumpster diving requires a throw that describes whether relevant objects were found and an analysis role to represent the evaluation of these finds. The same applies to baiting. A programming roll is needed to create the malware for the USB sticks and the second roll decides on the success of distributing these sticks.

\subsubsection{Points for Skills}

Due to the similarity of the sub-categories in terms of their required personal skills, they are divided into super-categories. To represent this connection in the game, bonus points should be distributed to all subclasses of a superclass if a certain number of invested points is reached or exceeded. Bonus points are awarded based on the investment (low: zero; medium: one; and high: two), in order to appropriately reward the distribution of points in a subclass. If a player is designed as a master of their craft in one superclass, their character should receive at least two of the high investment sub-categories and one medium investment sub-category. This gives them five bonus points for reaching the predetermined threshold of eleven points. A high investment is fixed at six points, which enables two bonus points in each corresponding sub-category. Proficiency is reached with five points. It makes sense to define the mean investment as three points, so that the same number of points distributed also results in the same number of bonus points. With that, characters can receive one bonus point in each corresponding sub-category.

According to the original idea, the effects of invested points are not additive. If the player wants to achieve the intended probability of 90\% with a random decision, they are forced to invest at least 15 points in a super-category to unlock a total of five bonus points on all associated subclasses. That being said, each character can be proficient in up to two sub-categories. If the player has invested a minimum of five points in more than two sub-categories, they must choose two of them. The reason for this is that, like in other TRPGs, the players should make a meaningful choice for character creation. In order to enable mixed classes and create more diversity, the skill points are limited to 21. Furthermore, a maximum of six invested points per sub-category is set to remain largely within the framework of the previously determined probabilities.

\subsubsection{Inventory of Proficient Skills}

Characters who are proficient in certain skills get an inventory item depending on the skill. Table~\ref{tab:skills} provides an overview of the inventory items. Aiming to make individual decisions easier for players and adding another element of strategy, players can lower the threshold of some decisions by up to four points by using the correct inventory item. Bribes, such as money, and custom IT gadgets, such as BadUSB tokens, can only be used once per scenario. Fake IDs and costumes can create more validity, especially with pretexting, or make a tailgating approach less risky. In some SE attacks, the inventory is even necessary to be successful, see the requirement of a server for a website. Smuggling of objects describes the unnoticed smuggling of small objects, such as a USB stick containing malware, into the target building. This procedure can be successful through security checkpoints, for example, by transporting a data carrier under the cast of a supposedly broken arm. Distraction is limited to the short-term distraction of non-player characters (NPCs), e.\,g., to escape without being noticed.

\subsubsection{Predefined Characters}

Two example character types, i.\,e., specialist (see Table~\ref{tab:specialists}) and jacks of all trades (see Table~\ref{tab:jacks}), help to establish new ones or can be used as they are. The respective subclasses selected as proficient are marked in gray.

\begin{table}[htb]
\centering
\caption{Specialists}
\label{tab:specialists}
\begin{tabular}{lll}
\toprule
\textbf{Wolf of Wallstreet} & \textbf{Nerd} & \textbf{Thief} \\ \midrule
\textcolor{darkgray}{Manipulation: 6} & \textcolor{darkgray}{Programming: 6} & \textcolor{darkgray}{Picking a lock: 6} \\
\textcolor{darkgray}{Occurrence: 6} & \textcolor{darkgray}{Analysis: 6} & \textcolor{darkgray}{Sleight of Hand: 6} \\
                    Adaptation: 3 & Design: 6 & Fitness: 3 \\
                    Fitness: 3 & Forging: 3 & Adaptation: 3 \\
                    Sleight of Hand: 3 & & Forging: 3 \\\bottomrule
\end{tabular}
\end{table}

\begin{table}[htb]
\centering
\caption{Jacks of all trades}
\label{tab:jacks}
\begin{tabular}{lll}
\toprule
\textbf{Spock} & \textbf{Con-Artist} & \textbf{Craftsman} \\ \midrule
\textcolor{darkgray}{Programming: 6} & \textcolor{darkgray}{Sleight of Hand: 6} & \textcolor{darkgray}{Forging: 6} \\
\textcolor{darkgray}{Manipulation: 6} & \textcolor{darkgray}{Manipulation: 6} & \textcolor{darkgray}{Picking the lock: 6} \\
                    Adaptation: 3 & Adaptation: 6 & Fitness: 6 \\
                    Analysis: 3 & Occurrence: 3 & Occurrence: 3 \\
                    Design: 3 & & \\\bottomrule
\end{tabular}
\end{table}

\subsubsection{Role of the Game Master}

The GM is responsible for all NPCs and describes the game environment, supported by the LEGO model. In addition, they make random decisions and judge the plausibility of the players' strategies, which, therefore, can also be rejected. In principle, the previously defined threshold value of 18 applies, but the GM can increase or decrease this by up to four. Should an attack fail, then the GM determines whether the consequences can be compensated. If this is not the case, the alertness of all NPCs and, hence, the thresholds depending on NPCs increase. This rise remains until the end of the game unless players find a way to ease the tension. As the threshold continues to grow with repeated failures, this could cause the scenario to fail and the players to have to start over.

\subsection{Scenarios}

Three scenarios, each with an increasing level of difficulty, for both SMEs and large organizations were designed as a start. In the following, we describe both introductory scenarios.

\subsubsection{SME: The Giveaway}

This scenario is based on the example provided in Section~\ref{sec:scenarios}. The players have to infiltrate the IT system of a new startup with eight employees, which basically has one large room, as shown in Section~\ref{sec:sme-play}. The players can get the following knowledge through reconnaissance:
\begin{enumerate}
\item the premises are small and insightful;
\item dumpsters contain many pizza boxes; and
\item the entrance is always locked.
\end{enumerate}

The second source of information is the social media account:

\begin{enumerate}[start=4]
\item the company has only recently been founded;
\item the employees are a group of friends; and
\item the company recently had a great success.
\end{enumerate}

Due to the small size and structure of the office, attackers are denied access methods such as tailgating or simply breaking in. The close relationship between the employees further restricts the possibilities. The love for pizza should give players the idea of impersonating the pizza delivery man. Based on the real-world scenario~\cite{pizza} (see Section~\ref{sec:scenarios}), a fake email with a link to the pizza website is an option to find a way into the building or deliver a freebie such as a USB device, which is infected with malware.

\subsubsection{Big Organization: Theft of Data}

The players have to steal the customer data of an insurance company, which can be described as a toxic workplace. The players can gain the knowledge that

\begin{enumerate}
\item due to hire and fire, there is a high fluctuation, employees socialize in circles, and the company is recruiting;
\item employees often do not lock their screens;
\item employees can be encouraged to talk;
\item some employees use the emergency exit to smoke in the car park, leaving the door unlocked;
\item an external company is hired to dispose of hardware at a low frequency to reduce costs;
\item access to server rooms is restricted;
\item computers of low-level employees are in a restricted subnet in contrast to the C-level (which resides on the second floor);
\item the C-level leaves daily at the same time for lunch in a nearby restaurant.
\end{enumerate}

The corresponding model is described in Section~\ref{sec:big-play}. The players get a first impression (see knowledge (1), (4), (8), and after some time or luck (5)) by simply observing the building. By talking to employees in the smoking area or looking at the company's website, the players can find out that they are recruiting (1). In addition, interaction with the employees leads to knowledge (3), (5), and maybe (2), (6), and (7). Consequently, there are several ways to proceed. Players can, e.\,g., pretend to be potential customers or candidates. Such an interview takes place on the second floor, where the computers to compromise the system are located. The players can steal and forge the key card of a C-level member or send a phishing email with the required malware from one of the computers.

\subsection{Models}

The LEGO models for SMEs and big organizations are based on the elements described in Section~\ref{sec:concept} and were created with Studio~2.0~\cite{studio}. The models and the scenarios increase in difficulty by including most elements of the models beforehand. Since only a full exploration of a floor leads to it being placed separately and visible to the players, an increasing difficulty from bottom to top is created. The consecutive models also reduce costs and time. We first outline the specified color, before we describe the models for SMEs and large organizations.

\subsubsection{Color Code}

In order to enable a simple and understandable structure of the models without generating high costs for special LEGO bricks, color coding is used for the interior design. Filing cabinets are bright yellow, trash is highlighted in green, and desks are white. In addition, the models for the large organization also color the doors to visualize countermeasures. Green doors are unlocked, blue doors are accessed by the majority of employees with keys or key cards, orange doors are restricted to a few employees, and red doors are limited to only specific employees.

\subsubsection{SME}
\label{sec:sme-play}

\begin{figure*}[!htbp]
  \centering
\begin{subfigure}[b]{0.3\textwidth}
\includegraphics[width=\textwidth]{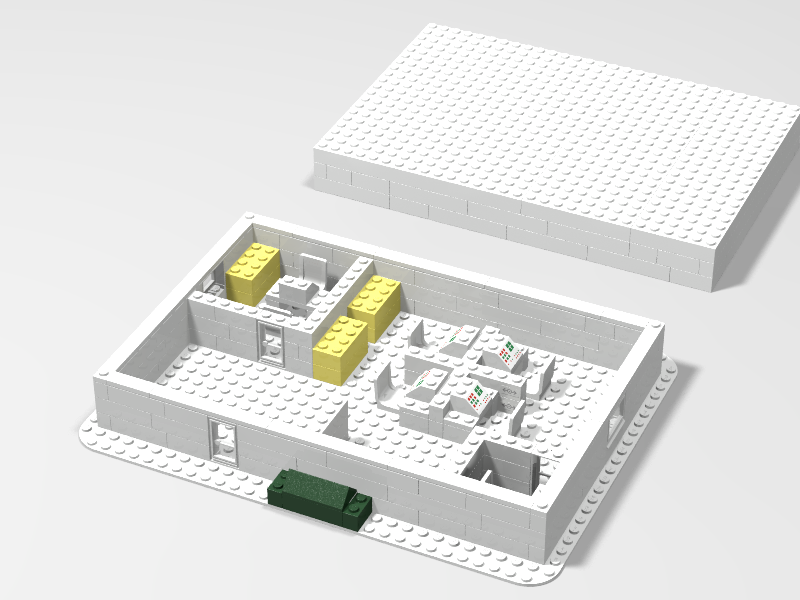}
\subcaption{Overview of the first model}\label{fig:pizza-overview}
\end{subfigure}
\begin{subfigure}[b]{0.3\textwidth}
\includegraphics[width=\textwidth]{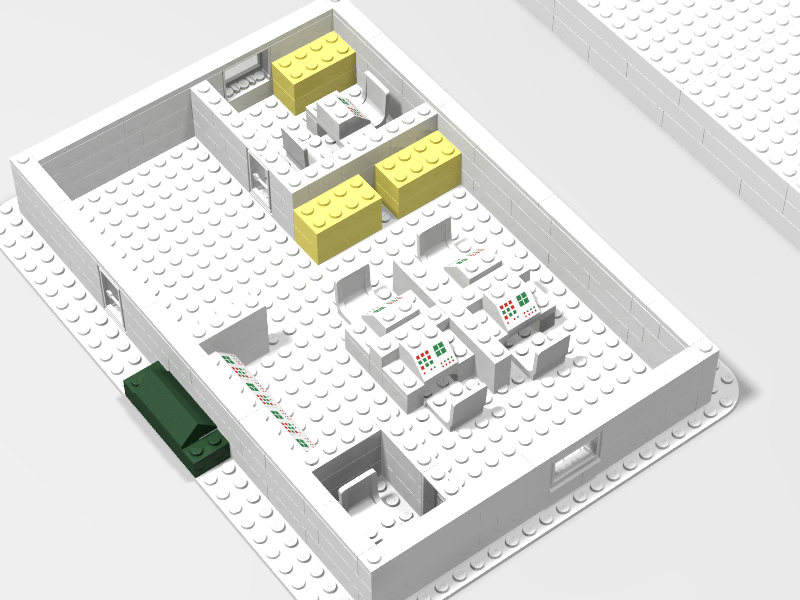}
  \subcaption{Close-up of the first model}\label{fig:pizza-eg}
\end{subfigure}
\begin{subfigure}[b]{0.3\textwidth}
 \includegraphics[width=\textwidth]{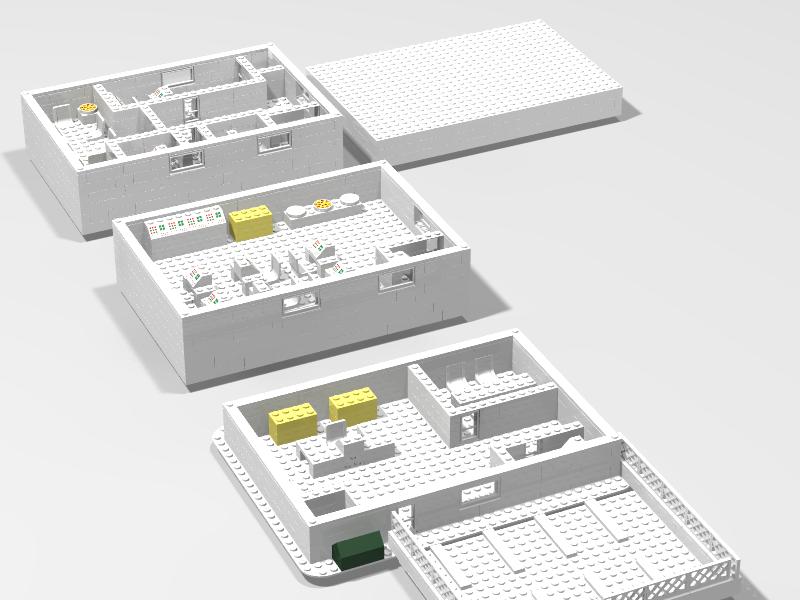}
 \subcaption{Overview of the third model}\label{fig:sme-third}
\end{subfigure}
  \caption{Model of the SME company building}
  \label{fig:sme}
\Description[Model of the SME company building]{Model of the SME company building with a close-up of the first model and an overview of the third model}
\end{figure*}

In Figure~\ref{fig:pizza-overview}, the compact building of the introduction scenario is shown. Figure~\ref{fig:pizza-eg} displays the amenities of the building. The garbage container at the entrance cannot be seen from the interior. Likewise, inside the building, it can only be roughly deduced that an undetected entry is not possible. The relevant electronics as well as the stored files are all compressed in one place, with the servers facing the dumpster. Last but not least, Figure~\ref{fig:sme-third} shows the third model, incorporating the models of the first and second scenarios, and, hence, consisting of three floors and a parking lot.

\subsubsection{Large Organization}
\label{sec:big-play}

\begin{figure*}[!htbp]
  \centering
\begin{subfigure}[b]{0.3\textwidth}
\includegraphics[width=\textwidth]{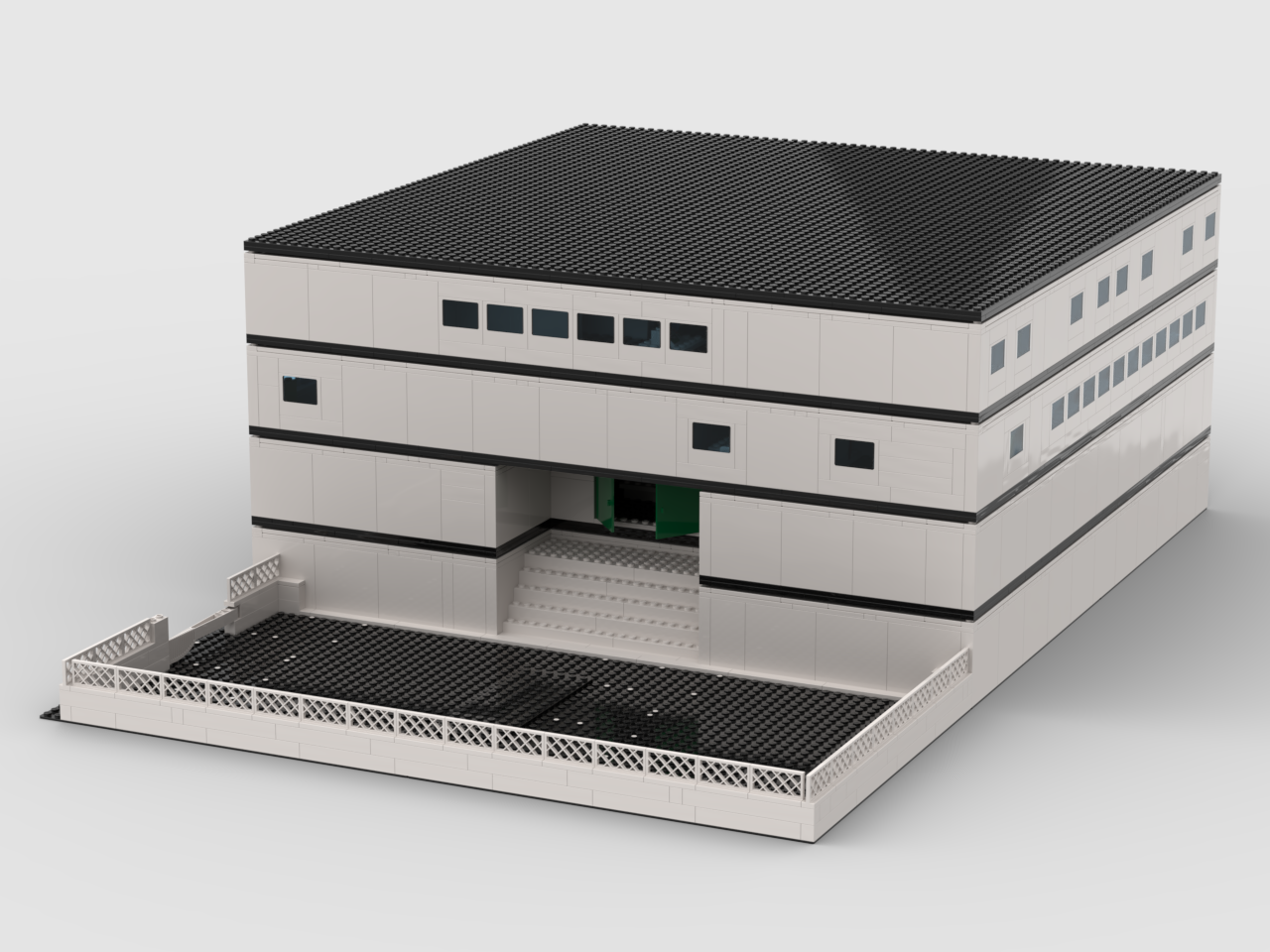}
\subcaption{Overview of the overall model}\label{fig:big-overview}
\end{subfigure}
\begin{subfigure}[b]{0.3\textwidth}
\includegraphics[width=\textwidth]{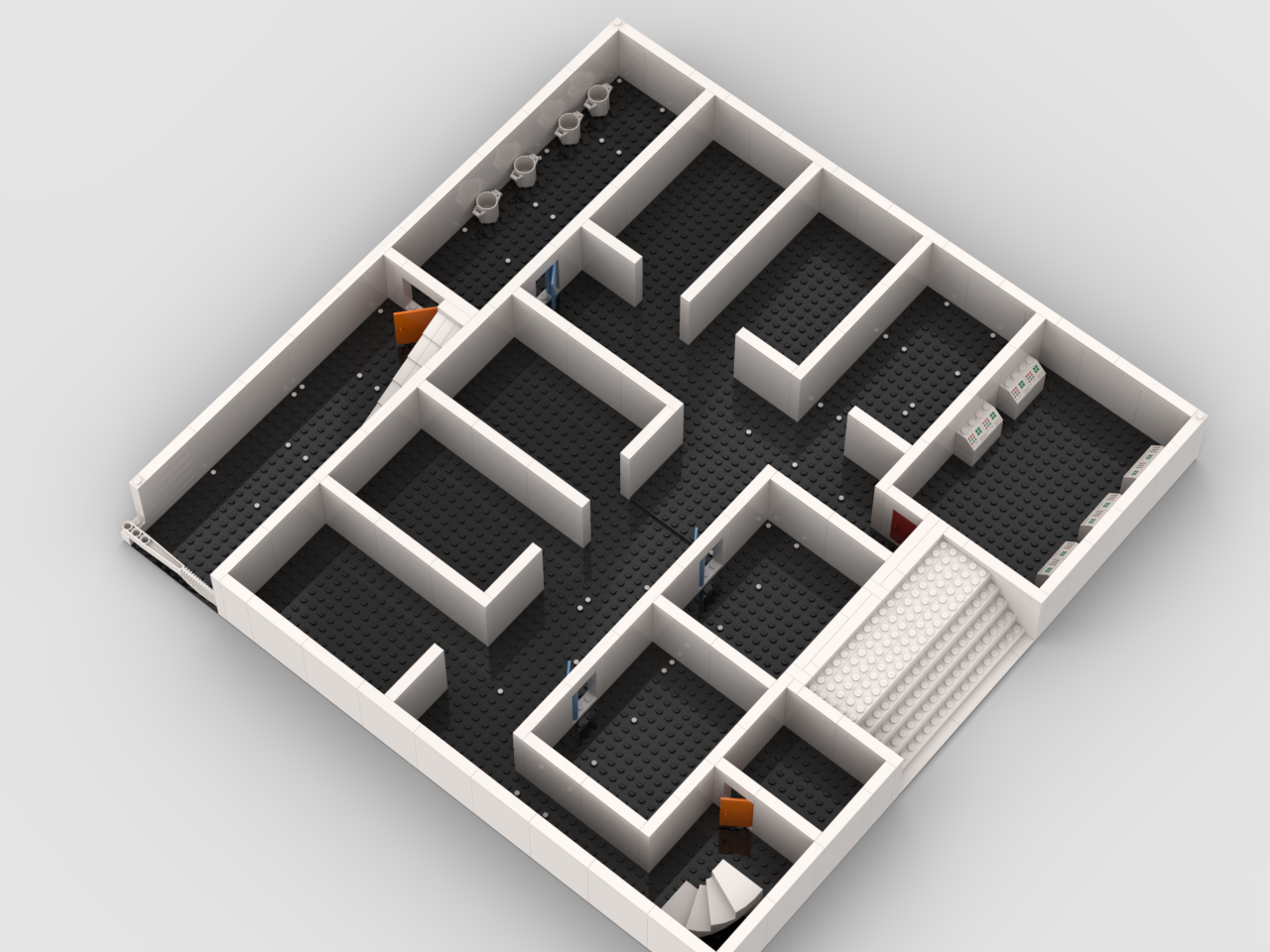}
  \subcaption{Close-up of the cellar}\label{fig:big-cellar}
\end{subfigure}
\begin{subfigure}[b]{0.3\textwidth}
 \includegraphics[width=\textwidth]{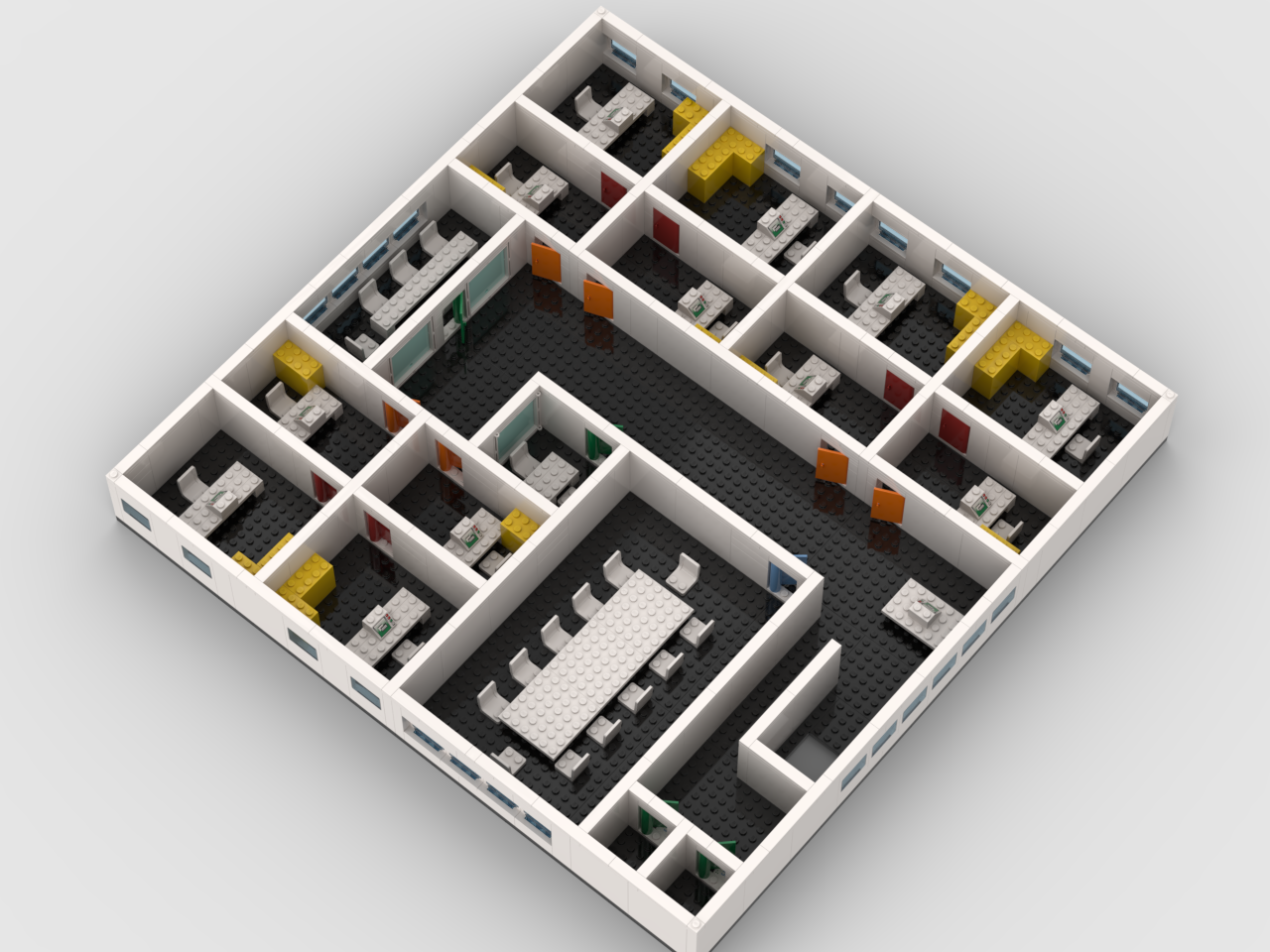}
 \subcaption{Close-up of the second floor}\label{fig:big2}
\end{subfigure}
  \caption{Model of the large organization's company building}
  \label{fig:big}
  \Description[Model of the large organization's company building]{Model of the large organization's company building with an overview, the cellar, and the second floor}
\end{figure*}

The large organization has the elements of ground, first, and second floors, a parking lot, and a cellar. An overview of the building without the cellar is given in Figure~\ref{fig:big-overview}. On the ground floor, there is a welcome desk opposite the entrance. In addition, the floor consists of customer advice, a server room, a storage room, and a staircase, which leads, among others, to the emergency exit. The bathrooms are located on the first floor, opposite the staircase. The main area of the floor is characterized by employees' workplaces and small meeting rooms. Further elements are the office of the head of the department, their secretariat, a balcony, and a staff kitchen. As can be seen in Figure~\ref{fig:big2}, the second floor mainly contains the offices of high-ranking employees and their associated secretariats. Along the main corridor are three meeting rooms of different sizes. The side passage next to the stairs leads to the sanitary facilities. Directly opposite the staircase is another workstation for a receptionist or security guard. The parking lot has an access restriction, which can be enforced more strictly depending on the scenario. The optional cellar offers another server room and various storage rooms, as shown in Figure~\ref{fig:big-cellar}. Since the ground floor has to be raised in order to make a cellar possible, the delivery access has to subsequently be rebuilt.

\section{Evaluation of TASEP}
\label{sec:evaluation}

In order to evaluate TASEP, we provided hands-on training for two different groups, playing the smallest scenario of the SME model.

\subsection{Study Design}

We recruited two groups of students with a minimum of three participants. The students were previously known to the selected GM. One group of students was technical-savvy with students of computer science in their masters. The other group of students was non-technical-savvy with studies in the humanities. An experienced GM with knowledge of SE was chosen. In addition, another expert served as an observer. The following methodology was applied.

\begin{enumerate}
\item We verified their knowledge with a multiple-choice test.
\item The participants were introduced to SE and the attack cycle.
\item The game principle of TASEP was explained.
\item TASEP was played.
\item The participants had to fill out feedback sheets and could provide oral feedback.
\item The game rounds and the feedback were evaluated. Here, the behavior of the participants and the course of the game rounds were compared. The insights of the GM and observer were taken into account. Last but not least, the feedback sheets were evaluated.
\end{enumerate}

The study was in compliance with the ethical boards and data protection laws and did not require specific acceptance. The study focused on the simplest model of SMEs for practical reasons. Hence, the results stand solely for this model and these participants.

\subsection{Multiple-Choice Test}

The participants had to answer 20 questions with 24 possible points to receive a baseline of existing knowledge about SE. The technical-savvy students had, although at different stages of their study, little variation in their results (2 points) and an average of 59\%. The difference among the non-technical-savvy students was a bit higher, with 5 points and an average of 40\% of the points. The higher variance, though, could be the result of the multiple-choice test. Nevertheless, it can be concluded that the second group had less basic knowledge of SE.

\subsection{Game Play}

\begin{figure}[!htbp]
  \centering
\includegraphics[width=\linewidth]{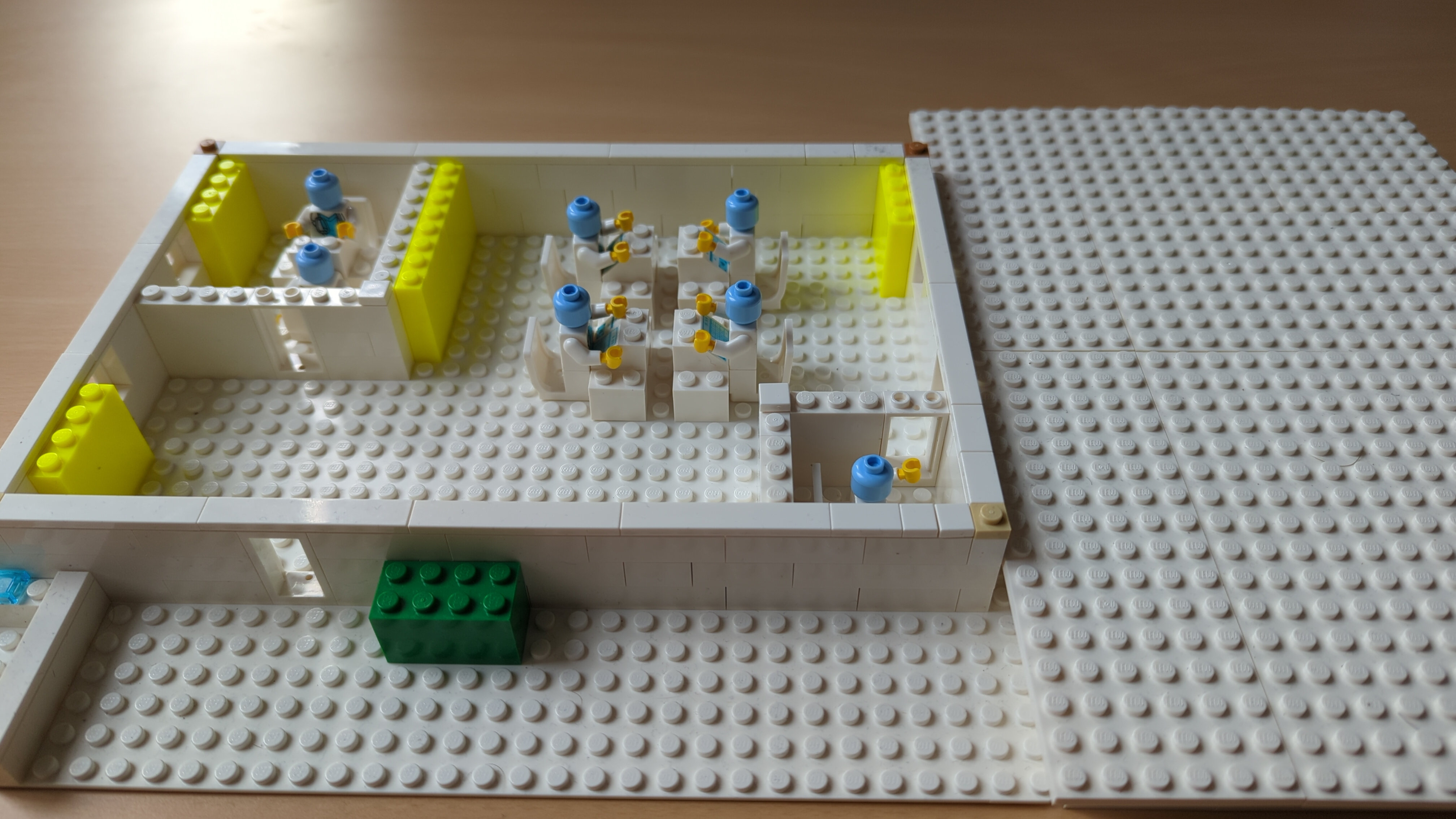}
  \caption{Picture of the built LEGO model}
  \label{fig:pic}
  \Description[Picture of the built LEGO model]{Picture of the built LEGO model}
\end{figure}

The players were introduced to the game, the goal of the game round, and the LEGO model (see Figure~\ref{fig:pic}).

\subsubsection{Technical-Savvy Students}

The technical-savvy students star\-ted their reconnaissance phase with an open source intelligence search on social media platforms, where they discovered, due to dice luck, the social structure of the startup, the love for pizza, and that the startup recently had a success. In the next step, the participants observed the building and found a nearby pizza restaurant. In order to get into the building, the group simulated a pizza restaurant and sent a phishing email to the startup with a decent discount on pizza. Contrasting the real-world pizza case, the participants used the address of the nearby pizza restaurant. To further proceed, they decided to pay the restaurant owner for participating in the penetration test. As a backup plan, the group again observed the building but failed to acquire the PIN code for the entrance through shoulder surfing. Hence, the participants focused on the pizza delivery and opted for a thief who impersonated a pizza delivery person. The impersonator took a large amount of pizza and asked to be helped and let in. In the following, the impersonator successfully plugged a prepared USB stick into an open USB port.

\subsubsection{Non-Technical-Savvy Students}

The non-technical-savvy students also started their reconnaissance phase with an open source intelligence search on social media platforms. In contrast to the other group, they only found out that the employees were friends. By dumpster diving, the group got to know that pizza is the favorite food of the employees. In the following, they opted to send a flyer with discounted pizza to the employees. In contrast to the first group, they used their own contact data. Additionally, the participants considered dropping USB sticks in front of the building. As it is situated on a main street, this option was discarded. After some discussions and brainstorming, they came up with the idea of USB loudspeakers as giveaways by the pizza restaurant and succeeded with this idea.

\subsection{Evaluation of the Game}

The rounds differed primarily in the focus of the two groups. While the technical-savvy students, who already had contact with the topic, were significantly more concerned with outsmarting the existing security measures, the humanities students were keener on direct contact with the victims. The latter resulted in a faster, but also more risky game round since the direct confrontation does not allow alternative ways to proceed. This risk could be due to misjudgment of the probabilities of success, but also because of the more confrontational personality of a socially active young adult compared to that of a technically oriented student. In addition, they were less inclined to plan out every detail of the procedure or to create the existing plan from scratch.

The evaluation sheets for the participants consist of generic questions about the game and the gameplay, with a Likert scale ranging from one (worst) to five (best). The scenario was rated as a good introduction to the subject of SE (avg. 4.66) with adequate difficulty (too easy: avg. 2.33). Also, the game design (avg. 4), the duration (avg. 4.66), the game flow (avg. 4.33), the game rules (avg. 4), the pre-defined characters (avg. 4), and the immersion (avg. 4) were rated positively. Although the attitude towards those attacks did not change during the game (avg. 2), a learning effect was noticeable. To achieve the higher goal of changing attitudes, the game will need to use all scenarios and, hence, be played either in a multiple-day block event or over the course of weeks.

Furthermore, the idea of expanding the rule book to include a level system for game characters was positively received (avg. 4.33). Thus, the future expansion of the game does not require any reference to sub-classes or new skills that a character can learn, but only a lower starting value of the skills, which can be increased with the successful completion of scenarios. Due to the mathematics on which the skill levels are based, it is advisable to consider the existing classes as the maximum value. In addition, guidance to establish and personalize characters (avg. 3) could be created.

\section{Discussion and Future Work}
\label{sec:discussion}

In order to evaluate the usage of TASEP in a course or for awareness campaigns, we first noticed that the attitude towards SE attacks had not changed. We assume that this is the case, as the participants regarded TASEP as a game with only one small scenario. This might be changed with several scenarios and a longer debriefing. We then discussed the idea of expanding the rule book, which was positively received. However, this would require an adaptation of the skill system and longer or more sessions. We only evaluated TASEP with two groups based on the smallest model and scenario. The other combinations still have to be tested. We observed differences in the strategies between both student groups. In future work, we want to explore differences in strategies based on study subject and attitude. Furthermore, we want to test TASEP with more participants and  measure knowledge retention and behavioral changes after TASEP sessions.

The immersion was rated positively. Hands-on challenges such as picking locks, copying access cards, and writing malware for USB drops could be incorporated to further improve immersion and introduce related content. These tasks could be highlighted with tangibles in the LEGO models. Although this would make the game more realistic, it also leads to longer gameplay. TASEP was explicitly designed as an offline board game. In order to allow remote participation, a hybrid mode could be implemented. However, not all hands-on challenges might be possible for online participants.

To make use of TASEP, at least one model, one scenario, and the game system are required. Either the model of the organization has to be created (at least partly) or some arbitrary house has to be used. Creating such a model and building it with LEGO requires time and effort, as the elements of the modular approach have to be determined, then the models have to be created, and then built. However, this effort is only needed once in the beginning. Scenarios can be chosen from this paper or adapted from real-world cases. For example, if an organization previously had an incident, this could be used as a scenario. In order to play such a scenario, all important elements have to be included in the model. Otherwise, there are almost no limitations. The game system itself can be taken from this paper. Since we introduced the game principle beforehand, no differences in the learning curve were noticeable. In addition, the principle of TASEP is simpler in comparison to D\&D, which helped the uptake.

\section{Conclusion}
\label{sec:conclusion}

A reverse real-world exercise is the most practical approach to raising awareness without pinpointing. However, this might not be possible everywhere. Hence, we proposed a concept for TASEP with models and gameplay based on the TRPG D\&D. This concept was put into practice with a skill set, individual skills, scenarios, and LEGO models for SMEs and big organizations. We successfully evaluated TASEP through two hands-on training sessions with technical-savvy and non-technical-savvy participants. Last but not least, we discussed our findings and possible future work. TASEP can be applied to introduce participants to the concept of SE attacks and raise awareness in the subsequent debriefing. Based on the evaluation, TASEP may result in a fundamental change of attitude, which was the goal of the training, but it requires more than one scenario to reach this goal.

\begin{acks}
This paper is part of project LIONS and is partly funded by dtec.bw -- Digitalization and Technology Research Center of the Bundeswehr. dtec.bw is funded by the European Union -- NextGenerationEU.
\end{acks}

%\clearpage
%\newpage
\bibliographystyle{ACM-Reference-Format}
\bibliography{se-tabletop}

\end{document}